\documentclass[twocolumn,showpacs,preprintnumbers,amsmath,amssymb]{revtex4}

\usepackage{graphicx}
\usepackage{dcolumn}
\usepackage{bm}

\begin{document}

\preprint{Proceedings of SPIE (2011)}

\title{Complex light with optical singularities induced by nanocomposites}

\author{Vlad V. Ponevchinsky}
\affiliation{Institute of Physics, NAS of Ukraine, 46 Prospect
Nauki, Kyiv 03650, Ukraine}

\author{Andrei I. Goncharuk}
\affiliation{Institute of Biocolloidal Chemistry named after F.
Ovcharenko,NAS of Ukraine, 42 Vernadskii Prosp., Kyiv 03142,
Ukraine, Tel. +380-44-424-03-78, Fax: +380-44-424-80-78}

\author{Sergei V. Naydenov}
\affiliation{Institute for Scintillation Materials of STC
"Institute for Single Crystals" of the National Academy of
Sciences of Ukraine, 60 Lenin Ave., 61001 Kharkov, Ukraine, Tel.
+380-57-341-03-58, Fax: +380-57-758-69-18}

\author{Longin N. Lisetski}
\affiliation{Institute for Scintillation Materials of STC
"Institute for Single Crystals" of the National Academy of
Sciences of Ukraine, 60 Lenin Ave., 61001 Kharkov, Ukraine, Tel.
+380-57-341-03-58, Fax: +380-57-758-69-18}

\author{Nikolai I. Lebovka}
\email{lebovka@gmail.com}
\affiliation{Institute of Biocolloidal Chemistry named after F.
Ovcharenko,NAS of Ukraine, 42 Vernadskii Prosp., Kyiv 03142,
Ukraine, Tel. +380-44-424-03-78, Fax: +380-44-424-80-78}%

\author{Marat S. Soskin}%
\email{marat.soskin@gmail.com}

\affiliation{%
Institute of Physics, NAS of Ukraine, 46 Prospect Nauki, Kyiv
03650, Ukraine}%

\date{\today}

\begin{abstract}
The nonocomposites on the base of long (5-10 $\mu$m, ${o}$-MWCNTs)
and short (2 $\mu$m, ${m}$-MWCNTs) multi-walled carbon nanotubes
(MWCNTs) hosted by nematic 5CB were investigated in details by
means of polarizing microscopy, studies of electrical conductivity
and electro-optical behaviour. The spontaneous self-organization
of MWCNTs was observed and investigated both theoretically and
experimentally. The efficiency of MWCNT aggregation in these
composites is controlled by strong, long ranged and highly
anisotropic van der Waals interactions, Brownian motion of
individual nanotubes and their aggregates. The simple Smoluchowski
approach was used for estimation of the half-time of aggregation.
It was shown that aggregation process includes two different
stages: fast, resulting in formation of loose aggregates
(${L}$-aggregates) and slow, resulting in formation of compacted
aggregates (${C}$-aggregates). Both ${L}$- and ${C}$- aggregates
possess extremely ramified fractal borders. Formation of the
percolation structures was observed for ${o}$-MWCNTs at
$C=C_p\approx$ 0.0025-0.05 \% wt and for $m$-MWCNTs at
$C=C_p\approx$ 0.005-0.25 \% wt. A physical model describing
formation of $C$-aggregates with captured 5CB molecules inside was
proposed. It shows good agreement with experimentally measured
characteristics. It was shown that MWCNTs strongly affect the
structural organization of LC molecules captured inside the MWCNT
skeleton and of interfacial LC layers in the vicinity of aggregate
borders. Moreover, the structure of the interfacial layer, as well
as its birefringence, drastically changed when the applied
electric voltage exceeded the Freedericksz threshold. Finally,
formation of the inversion walls between branches of the
neighbouring MWCNT aggregates was observed and discussed for the
first time.
\end{abstract}

\maketitle

\section{\label{sec:level1}INTRODUCTION AND MOTIVATION OF THE PROBLEM}

Complex light is an essential chapter of the modern optics. It is
especially actual for the light fields containing optical
singularities and special topological structures. Nanoscience and
nanotechnology are the hot spots now. Many promising composites on
the base of nanoparticles incorporated into continuous macro host
were recently proposed. The classical examples are colloidal
dispersions of nanoparticles in liquid crystals (LC) [1, 2].
Typically, concentration of nanofiller in a LC host is rather
small, which allows preserving the principal properties of the
host [3]. However, in many cases self-organisation of nanofiller
inside LC host provokes appearance of unusual nanostructures with
essentially new and nontrivial properties. Note that perturbation
of uniaxial nematic LC may result in birth of additional
topological structures up to singularities [3]. These effects may
be especially strong in nematics filled by nanoparticles. It was
recently demonstrated that propagation of laser beam through the
uniaxial nematic 5CB filled with multi-walled carbon nanotubes
(MWCNTs) generated the complex light and nano-particles originated
spontaneous birth of optical singularities in a LC host [4,5].

    The interesting data on the optical, electrophysical, and
thermodynamic properties of nematic 5CB filled with MWCNTs were
already reported [6-8].  Irreversible self-organisation of MWCNTs
results in formation of fractal aggregates [9] with extremely
ramified fractal borders [6,7]. An individual aggregate consists
of MWCNT skeleton with captured 5CB molecules inside. Moreover,
each aggregate perturbs the micron-sized interfacial LC layers
producing the random director orientations in the vicinity of the
aggregate. The hexagon dimensions in 5CB molecule match
excellently the hexagon structure of MWCNT [2] and theory [10]
predicts that 5CB molecules can be extra strongly anchored to the
side walls of nanotubes. The energy of interaction between 5CB and
surface of nanotubes is of the order of -2eV that is two orders of
magnitude higher than the thermal energy $kT$. Such strong
interactions may define unique optical and other physical
characteristics of the 5CB + MWCNT composite called legally
"scientific duo" [2]. Moreover, self-organisation and aggregation
of the particles with highly anisotropic shape in anisotropic
fluids is still far from full understanding and requires thorough
theoretical and fundamental studies [11].

The main aim of the present investigations was the study of
aggregation, electrical conductivity and electrooptical effects in
5CB filled by MWCNTs. The work discusses effects of spontaneous
self-aggregation and percolation, theoretical model of fractal
aggregation, capturing of LC molecules inside MWCNT skeleton and
interfacial LC layers in the vicinity of aggregates,
electrooptical data and electric field induced formation of
inversion walls.

\section{\label{sec:LII}
MATERIALS AND EXPERIMENTAL TECHNIQUES}

\subsection{\label{sec:L2.1} Liquid crystal }
5CB (4-pentyl-4'-cyanobiphenyl) was used as a nematic host. It was
obtained from Chemical Reagents Plant, Kharkov, Ukraine. The pure
5CB demonstrates the presence of a weakly first order isotropic to
nematic transition at $T_{ni}\approx 308-309$ K and a strongly
first order nematic to crystal transition at $T_{cn} = 295.5$ K.

\subsection{\label{sec:L2.2} Multiwalled carbon nanotubes}
The MWCNT preparation procedure is described in detail in [6]. It
is known that all the properties of carbon nanotubes depend
strongly on their aspect ratio [1, 2]. Two types of powdered
MWCNTs were investigated: (i) "long" $o$-MWCNTs with $5\div 10
\mu$m length (ii) and "short" $m$-MWCNTs with $\approx 2 \mu$m
length obtained by careful grinding of long MWCNTs in a mill. The
MWCNT aspect ratio was $\approx$ 500 and $\approx$ 200 for long
and short CNTs, respectively.

\subsection{\label{sec:L2.3} Preparation of MWCNT+5CB composites}
Before sonication, MWCNTs form dense blocks sized up to 50 $\mu$m,
which consist of nanotube bundles (tangles) [6]. Their mixture
with chemically pure 5CB was ultrasonicated (20 min) carefully up
using a UZDN-2T ultrasonic disperser to the apparently homogeneous
state.

\subsection{\label{sec:L2.4} Sandwich-type LC cells}
This mixture of 5CB + MWCNT was introduced into a sandwich-type LC
cell with 20 $\mu$m thicknesses [5] called sometimes "the flat
capillary" [12]. Polyimide alignment layers were unidirectional
rubbed for arrangement of the planar 5CB host texture [3]. The
transverse electric field could be applied to LC cell for
investigation of electrooptical effects including the Freedericksz
transition of the composite to the homeotropic orientation of 5CB
host.

\subsection{\label{sec:L2.5} Optical investigations}
Optical structures of nanocomposites were investigated and
documented by modern optical polarization microscope Olympus BX51
with colour high-resolution CCD camera and computer controlled
oven [13]. It allowed us to investigate the structure of 5CB +
MWCNT composites in crystal, LC and isotropic liquid phases. The
combined microscope objectives had prolonged focal length. Their
precise movement along the microscope axis ($\Delta z = 1 \mu$m)
allowed study of the optical polarization structures, including
topological structures around and between the aggregates of
nanocomposites inside the LC cell in various cross-sections [4-7].
The scattering of propagating laser beams was investigated using a
separate setup equipped by a He-Ne laser generating the lower
transverse mode, i. e. Gaussian beams.


\subsection{\label{sec:L2.7} Electrical conductivity measurements }
The electrical conductivity measurements were done by Instek 819
under the external voltage of 0.2 V and frequency of 10 kHz
applied to unaligned samples in a thick cell ($\approx$ 0.5 mm) in
the temperature range of 293-350 K (at 2 K/min scanning rate both
for heating and cooling). Each measurement was repeated, at least,
3 times for calculation of the mean values of experimental data.


\section{\label{sec:LIII}
RESULTS AND DISCUSSIONS}
\subsection{\label{sec:L3.1} Spontaneous self-aggregation and percolation threshold}

Freshly prepared 5CB+MWCNT composites undergo complicated
incubation processes comprehensively investigated recently by a
set of physical methods [6, 14]. The initial sonication resulted
in homogeneous dispersing of MWCNTs; however, after short period
of time a large quantity of spatially distributed aggregates were
formed. The transformation of homogeneous distribution of MWCNTs
into the segregated state is believed to occur owing to the
aggregation between different nanotubes. The efficiency of
aggregation is controlled by {\it(i)} strong and highly
anisotropic van der Waals interactions [1, 2] and {\it(ii)}
Brownian diffusion of individual nanotubes.

The simple Smoluchowski approach was applied to estimation of the
half-time of aggregation. In this approach, the MWCNTs were
modelled by cylindrical particles with huge aspect ratio
$r=l/d(r>>1)$, where $l$ is the length and $d$ is diameter of
MWCNT. Under assumption that cylindrical particles can aggregate
when the distance between them is of the order of their length
$l$, the half-time of aggregation $\theta$ can be estimated as
\begin{equation}\label{E01}
\theta=(4D \pi n l)^{-1},
\end{equation}
where $D$ is the diffusion coefficient and $n$ is the numerical
concentration of the particles. These values can be estimated as
\begin{equation}\label{E02}
D=kT \ln r/(3 \pi \eta r d)
\end{equation}
and
\begin{equation}\label{E03}
n=(\rho /\rho_ñ)C/V_ñ,
\end{equation}
respectively, where $kT$ is the thermal energy, $\eta$ is
viscosity,  $\rho/\rho_ñ$ is the ratio of densities of 5CB and
MWCNTs ($\rho/\rho_ñ\approx 0.5$), $C$ is the mass fraction of
MWCNTs in 5CB, and $V_c= \pi d^3r/4$ is the volume of a single
MWCNT.

Finally, the following relation can be obtained:
\begin{equation}\label{E04}
\theta\approx 3 \pi d^3 \eta/ (8kTC) (r/\ln r).
\end{equation}

Taking into account that $\eta\approx 0.1 $Pa.s (viscosity of
5CB), $d=$20 nm, $r$=500, $Ò$=300 K (temperature), we obtain
$\theta\approx 180$ s at $C=0.01$ \% wt and    $\theta\approx 18$
s at $C=0.1$ \% wt.

\begin{figure}[!htbp]
\centering\includegraphics*[width=0.4\textwidth]{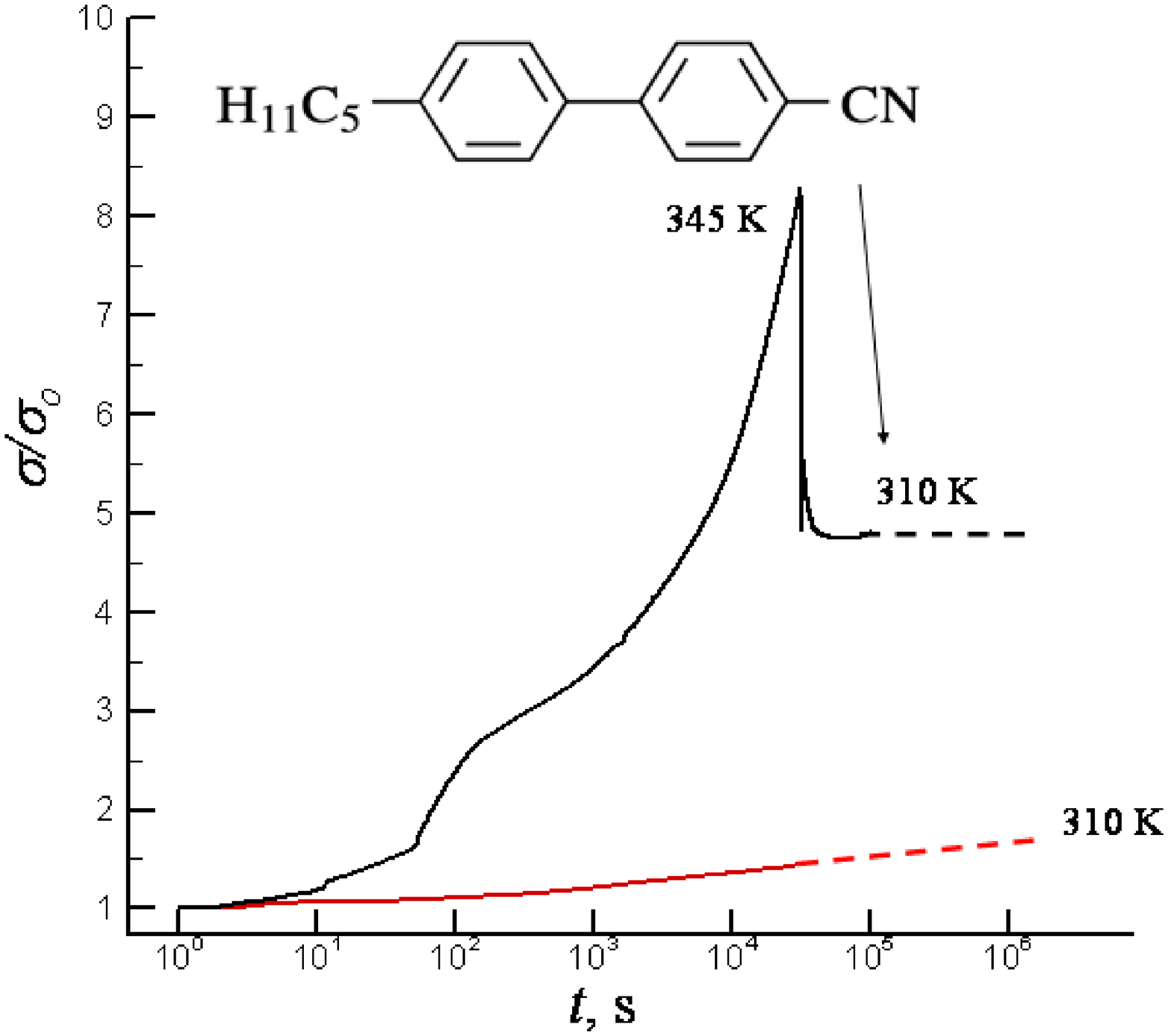}\hfill
\caption {\small Time dependencies of the relative electrical
conductivity $\sigma/\sigma_o$ ($\sigma_o$ is initial electrical
conductivity) in 5CB+o-MWCNTs composites (0.1 \% wt) incubated at
310 Ê and 345 Ê. The decrease of $\sigma/\sigma_o$ in a sample
incubated at T=345 Ê during its cooling to 310K reflects the
temperature effect on $\sigma$.}\label{fig:L01}
\end{figure}

The detailed investigations have shown that the structure of
nanocomposites and their physical characteristics attain stable
values only after several hours or days after sonication. It can
be explained by existence of the mechanism causing compacting of
the ramified, or loose, aggregates ($L$-aggregates) formed in the
"fresh" suspension into the more compacted aggregates
($C$-aggregate) formed in thermally incubated composite samples.
These structural changes were confirmed by the presence of changes
in electrical conductivity, which evidently reflects changes in
the microstructure of samples during the transformation of
$L$-aggregates to $C$-aggregates (Fig.1). The relative
conductivity $\sigma/\sigma_o$ ($\sigma_o$ is the initial
conductivity of the system after sonication) increased with time,
however, the observed effects were more pronounced at high
temperatures.

\begin{figure}[!htbp]
\centering\includegraphics*[width=0.4\textwidth]{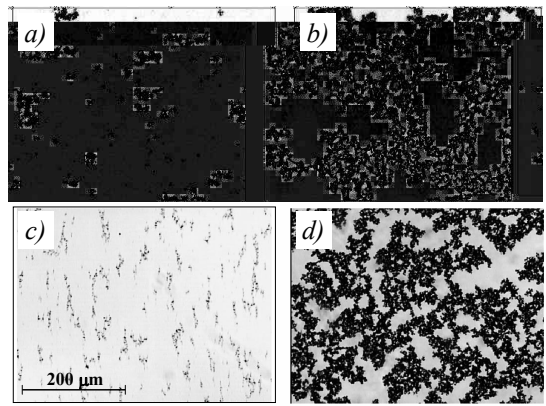}\hfill
\caption {\small Spontaneously self-aggregated clusters of "long"
($o$-MWCNT) (a, b) and "short" ($m$-MWCNT) (c, d) at various
concentrations of MWCNTs: a, c - onset of aggregation (0.0025 and
0.005 wt \%), b, d - percolation (0.025 and 0.25 wt \%).
}\label{fig:L02}
\end{figure}

The several (5-7) day incubation at room temperature resulted in stabilization
 of microstructure, optical and electrical conductivity characteristics,
and our experimental data presented below were obtained just for
such stable samples. Note that formation of $C$-aggregates with
irregular ramified borders [10] was visually observed by optical
microscopy investigations even at rather low concentrations of
MWCNTs (~0.01-0.1 \% wt). Such behaviour was observed both for
"long" ($o$-MWCNTs) and "short" ($m$-MWCNTs) nanotubes (Fig.2). In
both cases, the structural transformations were similar: from
small islands (Fig.2a, c) up to the percolation structure with
neighbouring clusters touching each other as their dimensions
increase with concentration of nanotubes (Fig.2b, d).

The cluster transverse dimensions start from few microns. Their
system expands over the whole LC cell when percolation structure
is reached. At this moment, the nematic 5CB host is broken into a
system of isolated "lakes" in 2D and sinuous "channels" in 3D.
Moreover, formation of the percolation structure is accompanied by
a drastic increase of electrical conductivity by 1-2 orders of
magnitude [4, 6, and 10]. It was observed that unmodified
nanotubes ($o$-MWCNTs) were forming the percolation networks at
$C=C_p\approx 0.025-0.05$ \% wt., whereas the modified nanotubes
($m$-MWCNTs) were forming the percolation networks at
$C=C_p\approx 0.25$\% wt. So, the percolation threshold
concentration was 5-10 time higher for $m$-MWNTs than for
$o$-MWNTs. Extremely small percolation threshold concentrations
are rather typical for such composites and can be explained by the
high aspect ratio ($r\approx $100-1000) of MWCNTs. Increase of the
percolation threshold for modified nanotubes ($m$-MWCNTs) ($C=C_p
\approx 0.1-0.25$\% wt.) possibly reflected the shorter length and
smaller aspect ratio $r$ of the modified nanotubes.

\subsection{\label{sec:L3.2} Theoretical model of fractal aggregation of MWCNTs in the nematic LC matrix: formation of $L$-aggregates }

The above-described formation of nanotube aggregates in the
nematic LC matrix can be better understood using the following
theoretical model. The MWCNTs of $L$-aggregates can be considered
as linear elements forming a compact skeleton of a certain regular
surface that bounds a certain confined space. We determine the
volume encompassed by the skeleton considering only one
characteristic spatial dimension of MWCNT- its linear length $l$ ,
which is fully justified if the aspect ratio $r >> 1$ ($r = l/d$,
where $d$ is the characteristic lateral dimension (diameter of
MWCNT).

From geometrical similarity considerations, the volume of a
cluster with characteristic linear dimension $l$ can be presented
as
\begin{equation}\label{E05}
V_{cl} = V_{cl}(l) = A_{cl}l^3,
\end{equation}
where $A_{cl}$ is a constant depending on the cluster geometry
(e.g., $A_{cl} = 1$ for cubic spatial structure, $A_{cl} =
\sqrt{2}/12$ for spatial structure formed by tetrahedrons, etc.).
Similarly, the total volume of the nanotubes comprising the
skeleton is $V_{nt}(l) = lB_{nt}g(d)$, where $B_{nt}$ is another
constant. The value of $B_{nt}$ depends upon cluster geometry
accounting for the number of nanotube edges comprising one cell of
the skeleton. Function $g(...)$ determines the volume of a single
nanotube in the cluster as a function of its characteristic
cross-section area. The form of $g(...)$ in the case of a regular
skeleton depends upon geometry of isolated nanotubes accounting
for the bends, fractures, deviations from cylindrical shape,
scatter of lateral dimensions, etc. In the case of fractal
geometry, $g(...)$ should "feel" how the nanotubes are connected
and should depend on the fractal "coastline" picture in the
direction normal to their predominant orientation. In a general
case
\begin{equation}\label{E05a}
g_f(x)\sim x^{d_f},
\end{equation}
where $d_f$  is the fractal dimension of the cluster and for
regular geometry $d_f$ = 2.

For regular skeleton geometry, the total volume of nanotubes in
the skeleton is
\begin{equation}\label{E06}
V_{nt}(l) = lB_{nt}g(d).
\end{equation}

If nanotubes are cylindrical,

\begin{equation}\label{E07}
g(d) = (\pi/4)d^2
\end{equation}
and
\begin{equation}\label{E08}
V_{nt} = B_{nt}l^3r_{-2}
\end{equation}
in terms of the nanotube length and aspect ratio. It can be
verified that $B_{nt} = 3\pi /2$ for tetrahedral skeleton cells;
$B_{nt} = 3\pi$ for cubic cells, etc...

Thus, the volume ratio of the cluster and the regular skeleton
(i.e., the ratio of the volume encompassed by the nanotube
skeleton and the total volume of the nanotubes involved) is
\begin{equation}\label{E09}
W_a = V_{cl}/V_{nt} = (A_{cl}/B_{nt})r^2 = Ar^2.
\end{equation}

With increase of aspect ratio $r$ (i.e., when nanotubes become
longer or thinner), $W_a$ grows rapidly, while the volume with
nematic molecules "captured" by the nanotube skeleton (the volume
of the "coat") increases much slower (not faster than linearly).
The coefficient $A$ depends on the specific skeleton geometry, and
its value can vary within $10^{-1} > A> 10^{-2}$ ($À =
1/3\pi\approx 0.1$ for cubic cells, $À \approx 0.025$ for
tetrahedral cells).

Thus, $W_a$ is  the ratio of the $L$-aggregate volume (i.e., the
volume of a loose skeleton formed by the nanotubes together with
"captured" molecules of the dispersion medium (nematic LC) both
"inside" the skeleton and in the nearest coordination layers) to
the total volume of nanotubes comprising the skeleton. If $C =
V_{nt}/V$ is the volume fraction (or mass fraction if the
difference between the densities of the dispersion medium and
nanotubes can be neglected) of nanotubes that were introduced into
the matrix, the value of $C_a = cW_a$ can be considered as
effective volume concentration of $L$-aggregates in the matrix
(solvent, dispersion medium).

In the case of fractal geometry, both the surface of the skeleton
formed by the nanotubes and the outer surface of $L$-aggregate
formed by the aggregated nanotubes and captured molecules of
dispersion phase will be of fractal character. Accounting for the
fact that $L$-aggregates are formed in orientationally ordered
matrix by orientationally ordered nanotubes, it is reasonable to
expect that these aggregates should be essentially oblate
(flattened). It leads us to a heuristic formula

\begin{equation}\label{Eq1}
W_a=V_{cl}/V_{nt}.
\end{equation}

Correspondingly, the effective concentration of aggregates in the
LC+MWCNT dispersion is

\begin{equation}\label{Eq2}
C_a=CW_a=Cr^{d_f}A_{cl}/B_{nt}=CAr^{d_f}.
\end{equation}

Thus, the effective concentration of $L$-aggregates at a given
initial concentration of nanotubes appears to be a universal
function of the aspect ratio, which is rather attractive. The
proportionality coefficient $À$ depends on specific geometry of
the nanotubes and the fractal skeleton structure formed by them;
it can be considered as constant for a given type of the nanotube
dispersions. As it was noted above, its value can be taken as
falling within 0.025 - 0.1.

This formula accounts at semi-empirical level for all main factors affecting aggregation of the nanotubes:
 the initial concentration of nanotubes $C$, their aspect ratio $r$, geometry of cluster formation $A$ and
 fractal dimensionality of the formed aggregates $d_f$.
 This formula can be easily verified experimentally.
 Thus, for two different MWCNT+LC dispersions prepared and studied under the same conditions
 but differing by certain parameters (e.g., aspect ratio $r$ and/or nanotube concentration $C$),
 the other measured characteristics should be related by equation (2).
 Preliminary verification can done using data from [13].
 In this paper, Fig.3 shows   values for three concentrations of the nanotubes
 (generally speaking, for three specific cases of $L$-aggregate formation).
 According to (1.3), at $r$ = 100 and $A$ = 0.1, we obtain for $C$ = 0.1 \% and $d_f$ = 1.81 the value of
 $W_a = 417$, i.e., the total volume of the formed $L$-aggregates with fractal dimensionality 1.81 will make
 $C_a$ = 41.7 \% from to total volume of the system (the remaining 58.3 \% is the volume occupied by orientationally
 ordered nematic molecules that remained "free", i.e., not captured by the aggregates).
 Accordingly, we get $C_a$ = 12 \% for $C$ = 0.01 \% and $d_f$ = 1.54.
 This is in a semi-quantitative agreement with results of microscopic observations:
 the aggregates are relatively few and visually occupy $\approx 10$ \% of the visible area in dispersions with 0.01 \% of MWCNTs,
 while in dispersions with 0.1 \% of MWCNTs there is much more aggregates, their size is larger,
 and they occupy about one half of the visible area.
 A similar effect is experimentally observed when we do not increase concentration of the MWCNT,
 but changes occur with time in a freshly dispersed sample (incubation phenomena).
 After several hours the fraction of area occupied by the aggregates substantially increases ([13], Fig.6).

Behavior of 5CB molecules inside the aggregates and in the
interfacial layers has features in common with the properties of
nematic molecules inside micropores, which is also a challenging
problem [19]. In this case, essential differences are also noted
in molecular mobility and physical properties of the "bulk" and
interfacially ordered portions of the nematic.

\subsection{\label{sec:L3.3} Capturing of LC molecules inside skeleton and interfacial LC layers in the vicinity of aggregates}

Analysis of the obtained data allows us to assume that formed
aggregates consist of a nanotube skeleton and a large number of
small 5CB molecules (sized 1.5.2.3 nm) captured inside this
skeleton. It reflects the effect of extra-strong anchoring of 5CB
to the side walls of MWCNTs. As a result, the average orientation
of 5CB molecules (nematic director) near the MWCNTs layers is
perturbed due to random orientation of the nanotube axes with
respect to the regular planar orientation of molecules in the 5CB
host. Due to the strong interaction between the MWCNTs and
neighbouring 5CB molecules, there appears a complex {\it
interfacial layer} of LC molecules near the surface of aggregates.
As a result, the MWCNT aggregates appear to be encapsulated by
such interfacial layers.

These layers with strongly perturbed LC directors are anchored to
the extremely ramified border of aggregate and extend to the bulk
5CB with unperturbed orientation of director. Formation of the
interfacial layers leads to new optical properties of 5CB + MWCNT
composites, resulting in creation of a new type of the complex
light. Indeed, one of the fundamental properties of LCs is
existence of elastic forces, which tend to orient the neighbouring
molecules in parallel to director [3]. These forces are expected
to be especially strong in the investigated composites due to
extra large energy of 5CB molecule anchoring to the nanotube. It
was shown that anchored 5CB molecules mostly retain their
positions in any phase, including isotropic liquid [4]. Quite
different is the structure of the outer part of the interfacial
layer, which smoothly transfers into the structure of the
unperturbed 5CB host and possesses microscale dimensions.

Strong elastic forces in the interfacial layers initiate
development of heterogeneous LC structures both along the
aggregate borders and in the transverse direction. This, in turn,
induces random local birefringence in the interfacial layer, which
manifests itself as bright spots on the dark background of
unperturbed 5CB in the microscopic image of the aggregated
composite observed through analyzer crossed to polarizer oriented
along alignment layers of the LC cell host (Fig.3a). It was
previously demonstrated that propagating laser beam gets scattered
or diffracted on the interfacial layers, thus acquiring
speckle-like structure full of optical vortices [4, 10]. Note that
direct microscopic observations using high-quality polarization
microscope [12] have confirmed existence of the microscale
interfacial layers and have shown that mean thickness of such
interfacial layers in 0.0025 wt \% composite was 1 $\mu$m [4].
Note that the size of 5CB molecules is only 1.5-2.3 nm, so, the
interfacial layers may include thousands of monomolecular 5CB
layers.

\subsection{\label{sec:L3.4} Electrooptical investigation of interfacial LC layers in the vicinity of aggregates}

The interesting responses of electrophysical properties to the
transverse external electric field applied to a sandwich-type LC
cell of thickness $d$ were already observed [3 - 6]. The 5CB
molecules are anchored strongly to the side walls of nanotubes and
can retain their position upon application of the electric field.
The birefringent structure of composites observed between the
crossed polarizer and analyzer shows that the structure of the
interfacial layer, as well as its birefringence, drastically
changes when the applied voltage $U$ approaches the Freedericksz
transition threshold (Fig.3).

\begin{figure}[!htbp]
\centering\includegraphics*[width=0.4\textwidth]{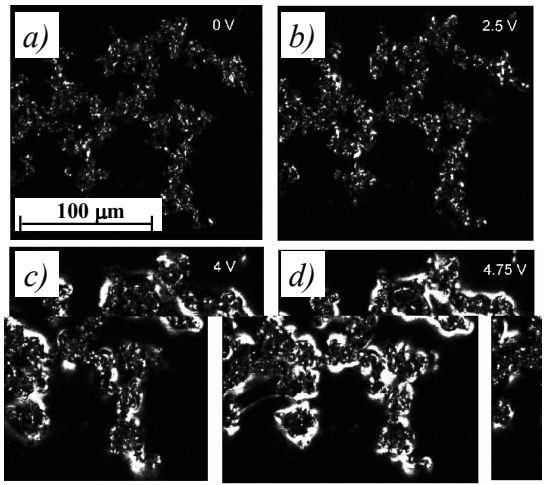}\hfill
\caption {\small Interfacial 5CB layer in the vicinity of the
aggregates of nanotubes (o-MWCNT, 0.005 wt \%) as a function of
the applied transverse electric field (analyzer A is crossed to
polarizer P oriented along the rubbing direction of the alignment
layers): a - without field, $U=0$;  b - $U=2.5$ V;  c - $U=4$ V; d
- $U=4.75$ V. The scale of black-and-white brightness is given on
the right. }\label{fig:L03}
\end{figure}

New isolated bright spots appear at $U$ = 2V when re-orientation
of 5CB molecules to the homeotropic alignment begins (compare
Fig.3a and Fig.3b). The mean thickness of interfacial layers in
0.0025 wt \% composite increases from 1 $\mu$m up to $\approx 4.5
\mu$m when nematic structure of 5CB undergoes the Freedericksz
transition from planar to homeotropic orientation under the
applied transverse electric field $\approx 2.5$ V [4]. Above the
Freedericksz threshold, a large portion of 5CB molecules in the
interfacial layers turn from initial planar to homeotropic
orientation, and the average thickness of the interfacial layer
increases (Fig.2c, at $U=$4 V) and reaches its maximal value at
$U=$ 4.5 V (Fig.3d).

However, the further electric field increase up to $U\approx$ 6 V
decreases the layer thickness to $\approx 2 \mu$m [4]. Further
increase of the applied field decreases the interfacial layer
thickness because most of 5CB molecules on the outer border of the
interfacial layer also re-orient to homeotropic arrangement. As a
result, the visible thickness of the interfacial layers around the
aggregates diminishes [4]. Note that detailed investigations have
shown the presence of quite different responses to the applied
field in the inner "lakes" and at the outer borders of MWCNT
clusters. These differences may be explained by different degree
of LC structure perturbation. The interfacial layers at the outer
borders are less perturbed and there exist smooth transition from
interfacial zone to unperturbed homeotropic state of 5CB.

\subsection{\label{sec:L3.5} Electric field induced formation of inversion walls (IW) between
branches of MWCNT aggregates}

Application of electric fields with voltage exceeding the
Freedericksz transition provokes also formation of singular
"channels" in the LC space between MWCNT aggregates. Figure 4
compares the microscopic images of the composite samples at $U$ =
0 V (a, b) and at $U$ = 3.5 V (c, d). The field-induced linear
topological structures resemble those known as inversion walls,
IWs [3, 16-20]. The IWs were first observed in a pure LC under the
transverse magnetic field [3]. They were obtained later by laser
irradiation of nematics in cells with small pre-tilt [20]. As in
the previously mentioned cases, in our experiments the IWs were
presented by three stripes, one bright central and two dark
lateral ones, which were observed between parallel polarizer and
analyzer: (Fig.4c, d).

\begin{figure}[!htbp]
\centering\includegraphics*[width=0.4\textwidth]{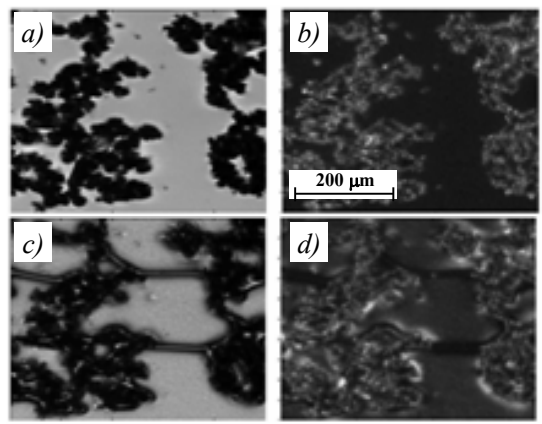}\hfill
\caption {\small Micro scale inverse walls in 5CB host between the
neighbouring aggregates of long carbon nanotubes (0.01 wt \%)
induced by the applied transverse electric field. The microscope
image of the actual aggregates is shown both without external
field, $U = 0$ V,  (a, b) and under the applied field, $U = 3.5$
V, (c, d). The polarizer and analyzer were parallel (a, c) and
crossed (b, d).}\label{fig:L04}
\end{figure}

Intensity distribution in IWs (Fig.5a) confirms that LC molecules
of the central stripe remain mainly planar-oriented, unlike the
auxiliary dark stripes with mainly homeotropic orientation of the
director, which is in full accordance with the previously reported
model (Fig.5b). Same as for IWs induced by magnetic fields in pure
LC [3], the total width of IW diminishes with increase of the
applied field (Fig.6). It can be speculated that spontaneous birth
of IWs between aggregates of MWCNTs is initiated by the electric
field-induced changes inside elastically strained overlapped
interfacial layers formed between different aggregates. Formation
of such IWs minimizes the internal stress in 5CB gaps between
MWCNT aggregates. The formation of IWs reflects self-organisation
in LC media that required free energy minimisation.

\begin{figure}[!htbp]
\centering\includegraphics*[width=0.4\textwidth]{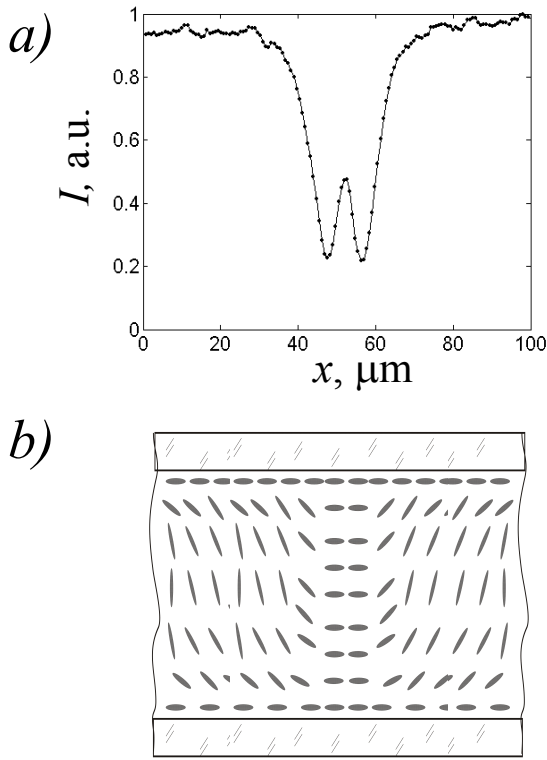}\hfill
\caption {\small Intensity distribution in the cross-section of
induced inversion walls between the aggregates for parallel
polarizer and analyzer (a). Schematic distribution of LC director
in a cross-section of the inverse walls (b).}\label{fig:L05}
\end{figure}

It was reported that IWs were produced on irregular borders of
nematic layer in a LC cell [16, 19]. Evidently, the observed IWs
were provoked by irregular fractal borders of MWCNT aggregates.
Moreover, the IWs appeared only between some branches of the
aggregates (Fig. 3). It seems reasonable to suggest that IWs
appear between branches that induce splay and bend distortions of
opposite sign (See [18], Fig.5c).

\begin{figure}[!htbp]
\centering\includegraphics*[width=0.4\textwidth]{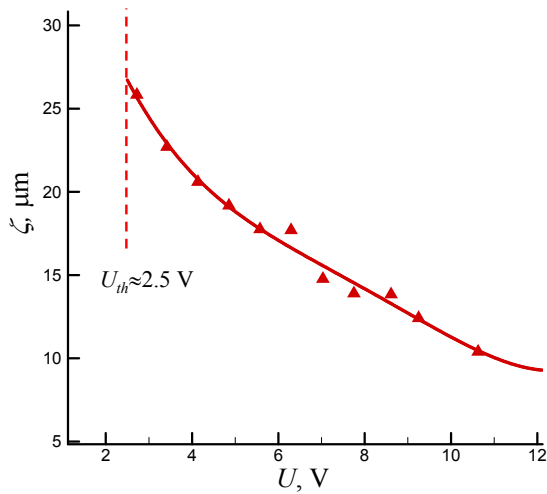}\hfill
\caption {\small Full width at half maximum  $\zeta$ of induced
inversion walls versus the applied field voltage
$U$.}\label{fig:L06}
\end{figure}

Figure 6 presents the full band width at half maximum $\zeta$ of
IW versus the applied field voltage $U$. The value of $\zeta$  was
maximal at $U=U_{th}\approx 2.5$ V and was decreasing with
decrease of $U$. Typically, no noticeable dependence of $\zeta$
upon the distance between different branches was observed.

\begin{figure}[!htbp]
\centering\includegraphics*[width=0.4\textwidth]{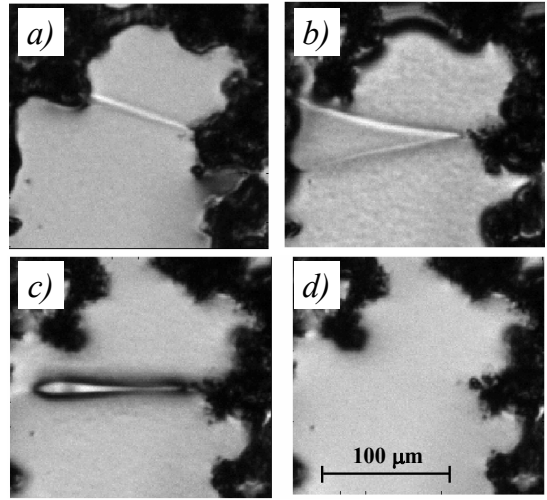}\hfill
\caption {\small Relaxation of the induced inversion walls when
the applied electric field $U$ = 9 V (a) is turned off for
different periods after turning the electric field off: 2s (b), 5s
(c), and 7s (d). }\label{fig:L07}
\end{figure}

Very elucidative is the transient evolution of the IW structure
when the electric field is turned off (Fig.7a-d). As an example,
in Fig. 7b (2s after turning electric field off), the left end of
the "channel" splits in two different walls that move to the
neighboring MWCNT branches. In Fig. 7c (5s after turning electric
field off), IW has the form of a closed loop; the similar
structure of IW was observed for pure LC under external magnetic
field [3]. Finally, the system relaxes to the initial state in 7s
after electric field is turned off. The observed effects were
completely reproducible when the electric field was repeatedly
turned on and turned off. This demonstrates that all the elastic
strengths involved into formation of IWs obey the Hook law.

Finally, we noticed that formation of such IWs is related with
some kind of self-organization minimizing elastic perturbations
introduced by the nanotubes and the effects of IW formation and
evolution in a LC composite system containing the nanotubes
deserve more detailed consideration in future, both from
experimental and theoretical points of view.

\section{\label{sec:LV}
CONCLUDING REMARKS}

The main task of our investigations was to study the influence of
carbon nanotubes on the micro and macro physical properties of
their LC nanocomposites. Many properties of 5CB + MWCNT composites
were thoroughly investigated by electrophysical methods. The major
findings may be formulated as follows:

1. Spontaneous self-organization of the nanotubes producing the
microscale aggregates in initially homogeneous dispersion was
observed and investigated both theoretically and experimentally.
The efficiency of aggregation is controlled by strong, long ranged
and highly anisotropic van der Waals interactions and the Brownian
diffusion of individual nanotubes. The simple Smoluchowski
approach was used for estimation of the half-time of aggregation;
it was shown that the process of aggregation includes the fast
stage resulting in formation of loosed aggregates ($L$-aggregates)
and the slow stage yielding compacted aggregates ($C$-aggregate).
The aggregates have ramified fractal borders; formation of the
percolation structures was observed at $C=C_p\approx 0.025-0.05$
\% wt for unmodified nanotubes (o-MWCNTs) and at $C=C_p\approx
0.1-0.25$\% wt. for the modified nanotubes ($m$-MWCNTs).

2. The proposed theoretical model of $L$-aggregation is in good
agreement with the experimental results. The microscale
interfacial layer of host 5CB molecules substantially affects the
optical and conductivity characteristics of the composites. Such
layer is extremely heterogeneous both along the fractal borders
and in cross-section. The inner borders of the layer are fixed to
the side walls of individual nanotubes due to extra strong
anchoring, in contrary to the outer borders, where orientation of
director approaches smoothly its orientation in the unperturbed
5CB host.

3. Transverse electrical field application to a sandwich-type LC
cell homeotropically orients the 5CB molecules, which leads to
increase of elastic strengths in the interfacial layer, its
thickness and induced birefringence. The inversion wall
topological structures were observed for the first time in LC+CNT
composites at threshold values of the applied field in the
vicinity of Freedericksz transition. Cross-section structure of
the inversion walls and their thinning with electrical field
increase were the same as those observed earlier in pure nematic.
The inversion walls were appearing between branches of some
neighbouring MWCNT aggregates. Full relaxation of inversion walls
was observed when the composite structure returned to its initial
state after switching off the applied field. It witnesses that all
the elastic strength changes obey the Hook law.

\section{\label{sec:LVI}
ACKNOWLEDGEMENT} Authors are thankful to Profs. Yuri Reznikov and
Victor Reshetnyak for useful consultations on LC physics, Dr.
Vasil' Nazarenko for polarization microscope Olympus, computer
controlled oven and useful LC discussions. This work was supported
in part by Projects 2.16.1.4, 2.16.1.7, ISTCU Project 4687).


\begin{thebibliography}{99}

\bibitem{Lagerwall2008}
Lagerwall, J. P. F., Scalia G."Carbon nanotubes in liquid
crystals". J. Mater. Chem. 18, 2890-2898 (2008).
\bibitem{Rahman2009}
Rahman M., Lee W. "Scientific duo of carbon nanotubes and nematic
liquid crystals", J. Phys. D: Appl. Phys., 42, (063001 (1-12)
(2009).
\bibitem{DeGennes1993}
De Gennes, P., Prost, J., "The physics of liquid crystals",
Clarendon Press, Oxford (1993).

\bibitem{Ponevchinsky2010a}
Ponevchinsky, V. V., Goncharuk, A. I., Vasil'ev, V. I., Lebovka,
N. I., and Soskin, M. S., "Cluster self- organization of nanotubes
in nematic phase: the percolation behaviour and appearance of
optical singularities", JETP Lett. 91, 239-242 (2010).

\bibitem{Ponevchinsky2010b}
Ponevchinsky, V., Goncharuk, A. I., Lebovka, N. I., Soskin, M. S.,
"Optical singularities induced in a nematic-cell by carbon
nanotubes", Proc. of SPIE Vol. 7613, 761306-1 (2010).

\bibitem{Minenko2010}
Minenko, S.S., Lisetski, L.N., Goncharuk, A.I., Lebovka, N.I.,
Ponevchinsky, V.V., and Soskin M.S., "Aggregates of  multiwalled
carbon nanotubes in nematic liquid crystal dispersions:
experimental evidence and a physical picture", Functional
Materials, 17, No.4, 454-459 (2010).

\bibitem{Lisetski2011}
7.  Lisetski, L. N., Minenko, S. S., Ponevchinsky, V. V, Soskin.,
M. S., Goncharuck, A. I., Lebovka, N. I., "Microstructure and
incubation processes in composite liquid cristalline material
(5CB) filled by muliwalled carbon nanotubes", Materials Science
and Engineering Technology/ Materialwissenschaft und
Werkstofftechnik, 42, p. xx-xx (2011) (accepted).

\bibitem{Ponevchinsky2011}
Ponevchinsky, V. V., Goncharuk, A. I., Minenko, S. S., Lisetski,
L. N., Lebovka, N. I., and Soskin, M. S. "Incubation Processes in
Nematic 5CB + Multi-Walled Carbon Nanotubes Composites: Induced
Optical Singularities and Inversion Walls, Percolation Phenomena",
Nonlin. Optics, Quant. Optics (submitted).

\bibitem{Feder1989}
Feder, J., "Fractals" (Plenum press, New York, London, 1989).

\bibitem{Kyung2007}
Park Kyung, Lee Seung, and Lee Young, "Anchoring a liquid crystal
molecule on a single-walled carbon nanotube", J. Chem. Phys. C,
111, 1620-1624 (2007).

\bibitem{Solomon2010}
Solomon, M. J., and Spicer, P. T., "Microstructural regimes of
colloidal rod suspension sol, gel, and glasses", Soft Matter, 6,
1391-1400 (2010).

\bibitem{Blinov1987}
Blinov, L. M., Katz, E. I., Sonin, A. A., Physics of thermotropic
liquid crystals", Uspechi fisicheskich nauk, 152, 449-477 (1987)
(in Russian).

\bibitem{olympusamerica}
http://www.olympusamerica.com

\bibitem{Goncharuk2009}
Goncharuk A. I., Lebovka N. I., Lisetski L. N., Minenko S. S.,
"Aggregation, percolation and phase transitions in nematic liquid
crystal EBBA doped with carbon nanotubes", J. Phys.D: Appl.Phys.
42, 165411-1-8 (2009).

\bibitem{Frunza2008}
Frunza, L., Frunza, S., Kosslick, H., and Shoenhals, A., "Phase
behaviour and molecular mobility of n-octycianobiphenyl confined
in molecular sieves: dependence on the pore size", Phys. Rev. E,
78, 051701 (2008).

\bibitem{Helfrich1968}
Helfrich, W., "Alignment-inversion walls in nematic liquid
crystals in the presence of a magnetic field", Phys. Rev. Lett.,
21, 1518-1521 (1968).

\bibitem{Demus1978}
Demus., D., and Richter, L., "Textures in Liquid Crystals", Frelag
Chemie, Weinheim, N. Y. (1978).

\bibitem{Janossy2001}
Janossy, I., and Prasad, S. K., "Optical generation of inversion
walls in nematic liquid crystals" Phys. Rev. E, 63, 041705 (2001).

\bibitem{Figueiredo1984}
Figueiredo Neto, A. M., Marnitot-Lagarde, Ph., and Durand G,
"Anisotropy induced director refraction inside inversion walls in
nematic liquid crystals", J. Physique Lett, 45, L-793 - L-798
(1984).

\bibitem{Cladis1987}
Cladis, P. E., van Saarloos, W., Finn, P. L., and Kortan, A. R.,
"Dynamics of Line defects in nematic liquid crystals" Phys. Rev.
Lett., 58, 222-225 (1987).

\end{thebibliography}

\end{document}